\documentstyle[prc,aps,axodraw,preprint]{revtex}
\begin{document}
\tightenlines
\title{One loop corrections to quantum hadrodynamics with vector mesons}
\author{Gary Pr\'{e}zeau}
\address{\small \sl Dept. of Physics, College of William and Mary,
  Williamsburg, Va 23187-8795}
\maketitle
\begin{abstract}
The renormalized elastic $\pi\pi$ scattering amplitude to one loop is
calculated in
the chiral limit in the $\sigma$ model and in a Quantum
Hadrodynamic model (QHD-III) with vector mesons.  It is argued that
QHD-III reduces to the linear $\sigma$
model in the limit that the vector
meson masses become large. 
The pion decay constant is also calculated to 1-loop in
the $\sigma$ model, and at tree level in QHD-III; it is shown
that the coefficient of the tree level term in the
scattering amplitude equals $F_\pi^{-2}$. 
The 1-loop
correction of $F_\pi$ in QHD-III violates strong isospin current
conservation.  Thus,
it is concluded that
QHD-III can, at best, only describe the strongly interacting nuclear
sector.

\end{abstract}
\narrowtext
\section{Introduction}

QCD is very successful at describing hadronic
interactions at high $Q^2$ where perturbation theory is applicable.
At low $Q^2$ however, non-perturbative methods must be used and
most quantitative predictions can not be extracted directly from QCD.
Models that are designed to describe low $Q^2$
hadronic interactions must be guided by the symmetries of QCD and
phenomenology.  It is a fact that QCD possesses global $SU(2)_L\times
SU(2)_R$ symmetry
and that large scale processes must involve meson loops because of
confinement.  Phenomenology also reveals the existence of conserved
vector and (partially) conserved axial vector currents.  QHD-III [1]
is a relativistic quantum field theory that
incorporates these features: in this
model, hadrons are the effective degrees of freedom and the vector
mesons that couple to these currents, the $\rho$ and the
$a_1$, are introduced as the
gauge bosons of a local $SU(2)_L\times SU(2)_R$ $\sigma-\omega$ model
with pions (see below).  The vector mesons are made massive via the
Higgs mechanism (although the Higgs scalars do not contribute to the
order considered in this paper) and the model is renormalizable.

A further motivation for QHD-III is the fact that simpler versions of
QHD based on $\{N; \sigma, \omega\}$ and $\{N; \sigma, \omega, \pi\}$
have
had significant phenomenological success [2]; hence, it is of interest
to see how
far this description can be extended.  A similar model with vector
mesons was developed in [3].  For
models with vector mesons in nonlinear chiral lagrangians, see
[4a,4b,4c,4d,4e].  For early work
on the subject, see [4f].
The consequences of the present model have not yet been
explored.  In the final analysis, the currents of QHD-III allow the
theoretical exploration of strongly interacting
systems and processes, while incorporating meson loop corrections.

This paper calculates to 1-loop the two simplest amplitudes in the
meson sector
of this model (the baryon sector is not included in this initial
investigation): $\pi\pi$ scattering and pion decay.  
First, the invariant $\pi\pi$ scattering amplitude ${\cal{M}}_{\pi\pi
\rightarrow \pi\pi}$ is calculated to
1-loop to ${\cal{O}}(g_\pi^4)$ and to ${\cal{O}}(g_\pi^2 g_\rho^2)$ in
the limit $m_\sigma^2\gg m_\rho^2 \gg s$ with $m_\pi=0$.  
To renormalize this scattering amplitude, the 
divergent parts of the
counterterms are extracted from the $\sigma$, 
$\sigma^2$
and $\sigma^3$ vertex functions, and it is shown that these
counterterms cancel all the divergences in ${\cal{M}}_{\pi\pi
\rightarrow \pi\pi}$ to ${\cal{O}}(g_\pi^4)$ and
to ${\cal{O}}(g_\pi^2 g_\rho^2)$, as well as the divergences in the
$\sigma$
4-point function as
expected in a renormalizable theory.  To 1-loop, it is also argued
that the gauge bosons decouple in the limit $m_\rho \rightarrow
\infty$.  Note that QHD-III reduces to the 
linear $\sigma$ model when $g_\rho=0$.  
Second, pion decay is analysed by looking at
the axial current matrix element $\langle 0| A_\mu^i | \pi^j \rangle$
in the $\sigma$
model to 1-loop; from it, the pion decay constant is identified.  To
renormalize the pion decay 
constant, the same counterterms evaluated to
${\cal{O}}(g_\pi^4)$ are used, and it is verified  that the
coefficient of the tree-level term in ${\cal{M}}_{\pi\pi \rightarrow
  \pi\pi}$ is the inverse square of the pion decay constant to this
order.  The matrix element $\langle 0| 
A_\mu^i | \pi^j \rangle$ is next considered in QHD-III so as to
identify the pion decay
constant in that model.  At tree-level, the pion decay constant of the
$\sigma$
model is replaced by $1/F_\pi^2 = 1/\sigma_\circ^2 +
g_\rho^2/m_\rho^2$.  To next order, it is found that the 1-loop
corrections
violate local current
conservation in this model; this matrix element to 1-loop is therefore
neither gauge-invariant 
nor renormalizable: to that order, {\it it is not an S-matrix element
of the theory}.
Thus the model can at best provide a phenomenological description
of the strongly interacting nuclear sector.
In the process of performing these calculations, the QHD-III
counterterm lagrangian with
coefficients $\{\delta_z,\delta_\mu,\delta_\lambda,\delta_{g_\rho},...\}$
is derived.

\section{QHD-III}

The model is constructed as follows: start with the $\sigma-\omega$
model with pions [2] (we use the conventions in [1]):
\begin{eqnarray}\label{so}
\cal{L}&=&\bar{\psi}[i\gamma^{\mu}(\partial_{\mu} + ig_V V_{\mu}) -
g_{\pi}(s+i\gamma_5  \tau \cdot \pi)]\psi +
\frac{1}{2}(\partial_{\mu}s \partial^{\mu}s + \partial_{\mu}
\pi\partial^{\mu} \pi )\nonumber \\
& & -\frac{1}{4}\lambda(s^2+ \pi^2 - v^2)^2 -
\frac{1}{4}F_{\mu\nu}F^{\mu \nu} + \frac{1}{2}m_V^2V_{\mu}V^{\mu} +
\epsilon s + {\cal{L}}_{ct},
\end{eqnarray}
where $\epsilon s$ is the chiral symmetry violating term and
${\cal{L}}_{ct}$ is the counterterm lagrangian.  The QHD-III lagrangian is
constructed as follows (details are given in [1]):  this
lagrangian is first made locally 
invariant under $SU(2)_L\times SU(2)_R$; 
this results in the appearance of the $ l_{\mu}$ and $
r_{\mu}$ gauge bosons coupled to conserved currents.  These bosons are
given mass through the
Higgs mechanism.  The mass matrix is then diagonalized and the $
l_{\mu}$ and $ r_{\mu}$ fields are replaced by the new
generalized coordinates, the $\rho_{\mu}$ and $a_{\mu}$.  The $O(4)$
symmetry is spontaneously broken by giving the
scalar field a vacuum expectation value ($s=\sigma_\circ - \sigma$
with $\sigma_\circ\equiv M/g_{\pi}$).  This in turn yields a bilinear
term in the lagrangian that must now be diagonalized by
redefining the
pion and $a_1$ field.  The end result for the meson sector is:
\begin{eqnarray}\label{diag}
\cal{L}_{\sigma\pi}&=&
\frac{1}{2}\left[\left(1-\frac{m_a^2}{m_\rho^2}\right)\partial_\mu\pi
\partial^\mu\pi +  (\frac{m_a}{m_\rho}\partial_{\mu} \pi +
g_{\rho}\sigma a_{\mu}
+ g_{\rho} \pi \times \rho_{\mu})^2\right]\nonumber\\ 
& & + \frac{1}{2}[(\partial_\mu\sigma -
g_{\rho}\frac{m_a}{m_\rho} \pi
\cdot a_{\mu})^2 - m_{\sigma}^2\sigma^2] - g_{\rho}^2 \sigma_\circ
a^{\mu}\cdot (\sigma a_{\mu}
+ \frac{m_a}{m_\rho} \pi \times \rho_{\mu})\nonumber\\
& &  +g_{\pi}
\frac{m_{\sigma}^2}{2M}
\sigma\left(\sigma^2 + \frac{m_a^2}{m_\rho^2}\pi^2\right) - 
g_{\pi}^2 \frac{m_{\sigma}^2}{8M^2}\left(\sigma^2 +
\frac{m_a^2}{m_\rho^2}\pi^2\right)^2 + {\cal{L}_{\sigma\pi}^\prime}.
\end{eqnarray}
In the above, the $a_1$ mass is given by
$m_a^2=m_\rho^2+g_\rho^2\sigma_\circ^2$ with $m_\rho$ the $\rho$ mass and
$g_\rho$ the $\rho-$nucleon coupling constant.  Equation (\ref{diag})
is referred to as the ``diagonalized lagrangian'' and
reduces to the linear $\sigma$ model when $g_\rho=0$. 
${\cal{L}_{\sigma\pi}^\prime}$ contains the ``new'' 
interactions that appear when the pion and $a_1$ fields are redefined;
to ${\cal{O}}(g_\rho^2)$ it is given by:
\begin{eqnarray}\label{new}
{\cal{L}_{\sigma\pi}^\prime}& = &\frac{g_\rho^2 \sigma_\circ}{m_\rho^2}
\left[ \sigma\partial_\mu\pi\cdot\partial^\mu\pi -
\pi\cdot\partial_\mu\pi\partial^\mu\sigma \right].
\end{eqnarray}

\section{results}

The scattering amplitude is considered first to ${\cal{O}}(g_\pi^4)$
by putting
$g_\rho=0$ and then to
${\cal{O}}(g_\pi^2g_\rho^2)$.  The loop
integrals are done using dimensional regularization in the metric
(+,-,-,-), and it is assumed
that
$m_\pi=\epsilon=0$ as well as $m_\sigma^2\gg m_\rho^2 \gg s,t,u$
where $s,t$ and $u$ are the Mandelstam variables (for discussions
regarding the $m_\sigma \rightarrow \infty$ limit of the linear $\sigma$
model, see [2,5,6,7]).  To
${\cal{O}}(g_\pi^4)$ in the $t$-channel,
the amplitude is ${\cal{M}}_{ac,bd}={\cal{M}}\delta_{ac}\delta_{bd}$
with: 
\begin{eqnarray}\label{pp1}
{\cal{M}}&=& \left[\beta t + \alpha_1t^2 + \alpha_2
(s^2+u^2) \right] \nonumber\\
& &+\frac{1}{F^4}\frac{1}{16\pi^2}\left[
-\frac{t^2}{2} \ln\!\frac{{\scriptscriptstyle{-}}t}{m_\sigma^2} -
\frac{1}{12}(3s^2+u^2-t^2)\ln\!\frac{{\scriptscriptstyle{-}}s}{m_\sigma^2}
-\frac{1}{12}(3u^2+s^2-t^2)\ln\!\frac{{\scriptscriptstyle{-}}u}{m_\sigma^2}
\right],\\ \nonumber \\ \label{f}
\beta &\equiv& \frac{1}{F^2}\left\{1-\delta_z -
\frac{3}{16\pi^2}\frac{m_\sigma^2}{F^2} 
\left[ \Gamma(\frac{\epsilon}{2}) + \ln 4\pi - \ln \frac{m_\sigma^2}{\mu^2} +
\frac{7}{6} \right] + 2(\delta_\mu +
\delta_\lambda)\right\}, \\ \label{alpha}
\alpha_1 &\equiv& -\frac{1}{F^4}\frac{1}{16\pi^2}\frac{49}{18} +
\frac{2}{m_\sigma^2 F^2} [2\delta_\mu +\delta_\lambda],
\hspace{3cm}\alpha_2\equiv -\frac{1}{F^4}\frac{1}{16\pi^2}\frac{2}{9}.
\end{eqnarray}
\begin{center}
\setlength{\unitlength}{1cm}
\begin{picture}(5.0,2.0)(0,0)
\GCirc(71,28){15}{0.5}
\DashArrowLine(28,8)(61,17){4}
\DashArrowLine(28,48)(61,40){4}
\DashArrowLine(80,17)(110,8){4}
\DashArrowLine(80,40)(110,48){4}
\Text(1,0.15)[]{$P_2,b$}
\Text(1,1.85)[]{$P_1,a$}
\Text(4,0.15)[]{$P_4,d$}
\Text(4,1.85)[]{$P_3,c$}
\end{picture}\\
Figure (1): {\sl $\pi\pi$ scattering.}
\end{center}
Here, $1/F^2\equiv
1/\sigma_\circ^2\equiv g_\pi^2/M^2$,
$\{\delta_z,\delta_\mu,\delta_\lambda\}$ are of ${\cal{O}}(g_\pi^2)$,
and $\mu$ parametrizes the
renormalization conditions.  The amplitude ${\cal{M}}$ does
not depend on $\mu$; the counterterms and the running coupling constant
conspire to ensure that.  The subscripts are defined in figure (1) and
twenty diagrams contributed to the above result. 
Note the first term in $\beta$ is just the 
tree-level amplitude.
Note also that the unitary corrections in the second line of equation
(\ref{pp1}) has the $\log$
structure in Mandelstam variables first derived by Lehmann
[8].  The divergent parts of
the counterterms are obtained by 
considering the $\sigma$ 1-point, 2-point and 3-point
functions.  The result to 1-loop is:
\begin{eqnarray}\label{ct}
\delta_\mu \doteq
-\frac{3}{2}\frac{g_\pi^2}{M^2}\frac{m_\sigma^2}
{16\pi^2}\Gamma(\frac{\epsilon}{2}); \hspace{2cm}\delta_\lambda \doteq
-2\delta_\mu,
\end{eqnarray}
with $\delta_z$ finite.  One can check by direct substitution that the
above counterterms determined from the scalar sector cancel the
divergences occuring in the $\pi\pi$ scattering amplitude.  This
verifies the counterterms calculated in [9].

For the scattering amplitude to ${\cal{O}}(g_\pi^2g_\rho^2)$, the
gauge bosons from QHD-III contribute another fifty diagrams.  The
corrections to
the parameters $\beta, \alpha_1$ and $\alpha_2$ are:
\begin{eqnarray}\label{pp2}
\delta\!\beta&=& \frac{g_\rho^2}{m_\rho^2}\left(1+
  \frac{1}{16\pi^2}\frac{m_\sigma^2}{F^2} \left\{-12 +
6\frac{m_\rho^2}{m_\sigma^2} \ln\!\frac{m_\sigma^2}{m_\rho^2}  +
9\frac{m_\rho^2}{m_\sigma^2} \left[\Gamma(\frac{\epsilon}{2}) + \ln\!
4\!\pi-
\ln\!\frac{m_\sigma^2}{\mu^2}+\frac{131}{162}\right]\right\}\right)
\nonumber\\
& &\hspace{6cm} + 2\frac{g_\rho^2}{m_\rho^2}\delta_{g_\rho}+
\left(8\frac{g_\rho^2}{m_\rho^2} - \frac{1}{\sigma_\circ^2}\right) 
\delta_z + \frac{2}{\sigma_\circ^2}(\delta_\mu +
\delta_\lambda), \\ \label{dalpha1}
\delta\!\alpha_1&=&-\frac{1}{F^2}\frac{g_\rho^2}
{m_\rho^2}\frac{6}{16\pi^2} 
\left[\Gamma(\frac{\epsilon}{2})+ \ln\!4\!\pi-
\ln\!\frac{m_\sigma^2}{\mu^2} -\frac{85}{108} \right]\nonumber
\\& &\hspace{5cm}-
8\frac{g_\rho^2}{m_\rho^2}\frac{\delta_{g_\rho}}{m_\sigma^2}
+2\frac{g_\rho^2}{m_\rho^2}\frac{\delta_z}{m_\sigma^2} 
- 4\frac{g_\rho^2}{m_\rho^2}\frac{\delta_\mu}{m_\sigma^2} +
  \frac{2}{m_\sigma^2\sigma_\circ^2} (2\delta_\mu
  +\delta_\lambda), \\ \label{dalpha2}
\delta\!\alpha_2&=&\frac{1}{F^2}\frac{g_\rho^2}{m_\rho^2}\frac{1}{16\pi^2}
\left[ \frac{26}{9} + \ln\!\frac{m_\sigma^2}{m_\rho^2}\right] +
2\frac{g_\rho^2}{m_\rho^2}\frac{\delta_{g_\rho}}{m_\sigma^2}.
\end{eqnarray}

Now, in equations (\ref{pp2})-(\ref{dalpha2}) and in the coefficient
of the $\log$ terms in
equation (\ref{pp1}), $1/F^2
\equiv 1/\sigma_\circ^2 +
g_\rho^2/m_\rho^2$ to ${\cal{O}}(g_\pi^2g_\rho^2)$ as required by
unitarity.  The
counterterms become:

\begin{eqnarray}\label{rhoct}
\delta_\mu\doteq\frac{3}{2}\left(g_\rho^2 -
\frac{m_\sigma^2}{\sigma_\circ^2}\right)
\frac{\Gamma(\frac{\epsilon}{2})}{16\pi^2};\hspace{2cm}\delta_\lambda
\doteq -2\delta_\mu;\hspace{2cm}\delta_z\doteq
6g_\rho^2\frac{\Gamma(\frac{\epsilon}{2})}{16\pi^2}.
\end{eqnarray}
The calculation is carried out in the unitary gauge.  It is shown that
the divergent contributions to amplitudes in the scalar sector are
gauge invariant; this is proved using the non-diagonalized lagrangian
which maintains explicit current conservation at each step.
Here, $\delta_{g_\rho}$ can be determined from $\rho$ decay, and it is
finite.
Note that $\delta_z$ has acquired a divergence.  Upon
substitution of equations (\ref{rhoct}) into equations
(\ref{pp2}) and (\ref{dalpha1}), it is found
that the amplitude is now \underline{finite} to
${\cal{O}}(g_\pi^2g_\rho^2)$; note that the $\sigma$ 4-point 
function is also made finite with these counterterms.  Hence, the
$\pi\pi$ scattering amplitude is now rid of all infinities (as is
the entire scalar sector).

Consider pion decay in the $\sigma$ model.  From the axial current,
$F_\pi$ is calculated to 1-loop to be:
\begin{eqnarray}
F_\pi=\sigma_\circ \left\{ 1+ \frac{\delta_z}{2} +
\frac{3}{32\pi^2}\frac{m_\sigma^2}{\sigma_\circ^2} 
\left[ \Gamma(\frac{\epsilon}{2}) + \ln 4\pi - \ln
\frac{m_\sigma^2}{\mu^2} +
\frac{7}{6} \right] - (\delta_\mu +
\delta_\lambda)\right\}.
\end{eqnarray}
First, note that the counterterms given in equations
(\ref{ct}) cancel the divergences in $F_\pi$.  Second,
notice that to ${\cal{O}}(g_\pi^4)$, $F_\pi^{-2}$ is
\underline{identical} to
$\beta$ given
in equation (\ref{f}).  This verifies a well known property of the
$\sigma$ model in the chiral limit [10].

From the axial current of QHD-III, the tree
level pion decay constant is found to be:
\begin{eqnarray}
F_\pi=\frac{m_\rho}{m_a}\sigma_\circ=
  \left(1+\frac{g_\rho^2}{m_\rho^2}\sigma_\circ^2\right)^{-\frac{1}{2}}
    \sigma_\circ,
\end{eqnarray}
or $1/F_\pi^2 = 1/\sigma_\circ^2 +
g_\rho^2/m_\rho^2$.  This result was first obtained by Gasiorowicz and
Geffen [3] (see also [4b]).  From
the first term in equation (\ref{pp2}), it is seen that the
relationship between the pion decay constant and the $\pi\pi$
scattering amplitude in the chiral limit is also verified in QHD-III
at tree-level. 

\section{discussion and conclusions}

Consider the ${\cal{O}}(g_\pi^2g_\rho^2)$ corrections to the $\pi\pi$
scattering
amplitude, equations (\ref{pp2})-(\ref{dalpha2}).  As $m_\rho
\rightarrow \infty$,  $\delta\!\alpha_1,\delta\!\alpha_2\rightarrow
0$.  This can be seen directly from equations (\ref{dalpha1}) and
(\ref{dalpha2}) by noticing that to this order, {\it i})
$\delta_{g_\rho}$ is finite in this limit and {\it ii}) the
${\cal{O}}(1/m_\sigma^2)$ terms are to be neglected
so that the only surviving contributions from the counterterms are
those
of the $\sigma$ model in equations (\ref{ct}).
Note also that the tree-level correction of $\beta$ [the first
term of equation (\ref{pp2})] goes to zero in that limit. 
However, in this amplitude, the limit $m_\rho \rightarrow \infty$
must be taken
without violating the constraint $m_\sigma^2/m_\rho^2 \gg 1$ but
finite.  This
constraint implies the appearance of a new (quadratic) divergence in
$\beta$ proportional to $m_\sigma^2$.  The $\ln
m_\sigma^2/\mu^2$ divergences in $\beta$ and $\delta\!\beta$ are
already absorbed in the renormalization of
the parameters of the lagrangian and need not be considered further.
The new quadratic
divergence in $\beta$ renormalizes
the pion decay constant to 1-loop in exactly the same fashion; in
[6,7], it
is shown quite generally that the
linear $\sigma$ model reduces
to the non-linear $\sigma$ model in the limit $m_\sigma\rightarrow
\infty$ and that quadratic divergences have no observable effect to
1-loop.
In $\delta\!\beta$, note that the 1-loop corrections are finite
constants in the heavy mass limit subject to the above constraint.
They are thus negligible with respect to the quadratic
contributions in $\beta$.  Hence, the gauge boson contribution to the
scattering amplitude
becomes negligible in the heavy mass limit: the QHD-III scattering
amplitude reduces to
the $\sigma$ model amplitude in the limit $m_\rho
\rightarrow \infty$ with $m_\sigma^2/m_\rho^2 \gg 1$ but finite.
 \footnote{Since the 1-loop corrections in $\beta$ and
$\delta\!\beta$ become
larger then the tree-level terms in the limit $m_\rho,m_\sigma
\rightarrow
\infty$, taking these limits in our scattering amplitude
is questionable since we used perturbation theory to obtain our
result. However, 4-point functions can be used to construct effective
lagrangians in the heavy mass limit [6], and our scattering amplitude
would have the same structure as that effective lagrangian.} 

The decoupling of the $a_1$ and the $\rho$ can be understood more
generally as follows:
from the lagrangian given in equation (\ref{diag}) and the definition
of $m_a$,
it is seen that the gauge bosons decouple from the
$\pi$ and the $\sigma$ when the gauge fields are rescaled according to
$\rho_\mu=\rho_\mu^\prime/m_\rho$ and $a_\mu=a_\mu^\prime/m_\rho$
and the
limit $m_\rho \rightarrow \infty$ is taken; this procedure results in
the lagrangian of the
linear $\sigma$ model (for a similar discussion in the linear $\sigma$
model
see [2]).  This suppression in inverse powers of the mass is
essentially the decoupling theorem [11].

Now consider pion decay in QHD-III.  When the 1-loop correction of the
matrix element $\langle 0| A_\mu^i |
\pi^j \rangle$ is evaluated using the axial current derived from the
QHD-III lagrangian, it is found that the counterterms given in equations
(\ref{rhoct}) {\it fail} to cancel all of the divergences;
it is also explicitly found that this matrix element is not gauge
invariant to 1-loop in QHD-III.  This violation of current conservation
occurs because the pion acquires a strong isospin charge when the
vector mesons are introduced as gauge bosons as is the case in
QHD-III; hence, in
a process which destroys strong isospin charge such as pion decay,
vector and axial-vector isospin current conservation is
violated,\footnote{Electromagnetic charge also disappears in pion
  decay.  Of course, it is carried off in the lepton sector.  The
author knows of
  no simple
  way to fix up strong vector and axial vector isospin current
  conservation in this model for pion decay.} and that process is not
an S-matrix
element of QHD-III.

In {\it summary}, the renormalized $\pi\pi$ scattering amplitude has
been
calculated to 1-loop in the $\sigma$ model and in QHD-III in the
chiral limit for small external momenta with respect to the masses.
The pion
decay constant has been calculated to 1-loop in the $\sigma$ model and
it is shown explicitly that $F_\pi^{-2}=\beta$.  It is also shown that
this
relation holds at tree-level in QHD-III.  It is
argued that the gauge bosons decouple from the pion and the $\sigma$ in
the heavy mass limit.  The 1-loop correction to $F_\pi$ in
QHD-III is seen to violate strong isospin current conservation; thus,
the model can at best describe the strongly interacting nuclear sector.

The author would like to thank J.D. Walecka for his guidance and
useful discussions during the completion of this work.  He
would also like to thank
C. Carlson, J. Goity and M. Sher for very worthwhile discussions.
This work was supported under DOE Grant No DE-FGO2-97ER41023.

\section{references}

[1] B.D. Serot J.D. Walecka, Acta Phys. Pol. {\bf B23} (1992) 655

[2] J.D. Walecka, {\it Theoretical Nuclear and Subnuclear Physics},
Oxford (1995).

[3] S. Gasiorowicz, D. Geffen, Rev. Mod. Phys. {\bf 41} (1969) 531.

[4] (a) Fayyazuddin, Riazuddin, Phys. Rev. {\bf D36} (1987) 2768;
(b) U. Mei$\beta$ner, Phys.

\hspace{16pt}Rept. {\bf 161} (1988) 213; (c) G. Ecker,
J. Gasser, A. Pich, E. de Rafael, Nucl. Phys. 

\hspace{16pt}{\bf B321}
(1989) 311; (d) G. Ecker, J. Gasser, H. Leutwyler, A. Pich, E. de Rafael,
Phys. 

\hspace{16pt}Lett. {\bf B223} (1989) 425; (e) M. Bando, T. Kugo, K. Yamawaki,
Nucl. Phys. {\bf B259} 

\hspace{16pt}(1985) 493; (f) J. Schwinger, Phys. Lett. {\bf 24B}
(1967) 473.

[5] J. Gasser, H. Leutwyler, Ann. Phys. {\bf 158} (1984) 42.

[6] R. Akhoury, York-Peng Yao, Phys. Rev. {\bf D25} (1982) 3361.

[7] T. Appelquist, C. Bernard, Phys. Rev. {\bf D23} (1981) 425.

[8] H. Lehmann, Phys. Lett. {\bf 41B}, (1972) 529.

[9] M.E. Peskin, D.V. Schroeder, {\it An Introduction to Quantum
  Field Theory}, Addison-

\hspace{16pt}Wesley (1995) Chap. 11.

[10] S. Weinberg, {\it The Quantum Theory of Fields II}, Cambridge
(1996) 200.

[11] T. Appelquist, J. Carazzone, Phys. Rev. {\bf D11} (1975) 2856.

\end{document}